\RequirePackage{fix-cm}
\documentclass[twocolumn,epjc3]{svjour3}
\smartqed  
\RequirePackage{graphicx}
\RequirePackage{mathptmx}      

\usepackage{graphicx} 
\usepackage{dcolumn}  
\usepackage{colordvi}
\usepackage{color}
\usepackage{epstopdf}
\usepackage{amssymb}
\usepackage{url}
\graphicspath{ps}
\usepackage{hyperref}
\usepackage{tabularx}
\usepackage{amsmath}
\usepackage{multirow}
\usepackage[title,titletoc]{appendix}
\renewcommand{\thesection}{\arabic{section}}
\renewcommand{\thesubsection}{\thesection.\arabic{subsection}}
\renewcommand{\thesubsubsection}{\thesubsection.\arabic{subsubsection}}

\makeatletter
\renewcommand*\fps@figure{p}
\@fpsep\textheight
\makeatother

\RequirePackage[numbers,sort&compress]{natbib}

\RequirePackage{doi}
\RequirePackage{xspace}
\RequirePackage{siunitx}
\RequirePackage{bm} 
\RequirePackage{xfrac} 
\RequirePackage{relsize} 

\newcommand{\zprime}{\ensuremath{Z\text{'}}\xspace}
\newcommand{\ptThrust}       {\ensuremath{\text{p}^*_{\text{t}, \text{thrust}}{}(\mu)}\xspace}
\newcommand{\ptMaxWrtMin}    {\ensuremath{\text{p}^*_{\text{t}, \mu_{\text{min}}}(\mu_{\text{max}})}\xspace}
\newcommand{\plMaxWrtMin}    {\ensuremath{\text{p}^*_{\text{l}, \mu_{\text{min}}}(\mu_{\text{max}})}\xspace}
\newcommand{\ptMumu}         {\ensuremath{\text{p}^*_{\text t} (\mu^{+}\!\mu^{-})}\xspace}

\newcommand{\tev}{\ensuremath{\mathrm{\,Te\kern -0.1em V}}\xspace}
\newcommand{\gev}{\ensuremath{\mathrm{\,Ge\kern -0.1em V}}\xspace}
\newcommand{\mev}{\ensuremath{\mathrm{\,Me\kern -0.1em V}}\xspace}
\newcommand{\kev}{\ensuremath{\mathrm{\,ke\kern -0.1em V}}\xspace}
\newcommand{\ev}{\ensuremath{\mathrm{\,e\kern -0.1em V}}\xspace}
\newcommand{\gevc}{\ensuremath{{\mathrm{\,Ge\kern -0.1em V\!/}c}}\xspace}
\newcommand{\mevc}{\ensuremath{{\mathrm{\,Me\kern -0.1em V\!/}c}}\xspace}
\newcommand{\gevcc}{\ensuremath{{\mathrm{\,Ge\kern -0.1em V\!/}c^2}}\xspace}
\newcommand{\mevcc}{\ensuremath{{\mathrm{\,Me\kern -0.1em V\!/}c^2}}\xspace}

\newcommand{\invfb}   {\ensuremath{\mbox{\,fb}^{-1}}\xspace}

\journalname{Eur. Phys. J. C}
\begin{document}

\title{Punzi-loss}
\subtitle{A non-differentiable metric approximation for sensitivity optimisation in the search for new particles}


\author{F.~Abudin\'en\thanksref{addr13}, M.~Bertemes\thanksref{addr15}, S.~Bilokin\thanksref{addr17}, M.~Campajola\thanksref{addr3b,addr8}, G.~Casarosa\thanksref{addr3,addr10}, S.~Cunliffe\thanksref{addr2}, L.~Corona\thanksref{addr3,addr10}, M.~De~Nuccio\thanksref{addr2},
G.~De~Pietro\thanksref{addr11}, S.~Dey\thanksref{addr19}, M.~Eliachevitch\thanksref{addr20}, P. Feichtinger\thanksref{e1,addr15}, T.~Ferber\thanksref{addr14},  J.~Gemmler\thanksref{addr14}, P.~Goldenzweig\thanksref{addr14}, A.~Gottmann\thanksref{addr14}, E.~Graziani\thanksref{addr11}, H. Haigh\thanksref{addr15}, M.~Hohmann\thanksref{addr21}, T.~Humair\thanksref{addr18}, G. Inguglia\thanksref{addr15}, J. Kahn\thanksref{addr6}, T.~Keck\thanksref{addr14}, I.~Komarov\thanksref{addr2}, J.-F.~Krohn\thanksref{addr21}, T.~Kuhr\thanksref{addr17}, S.~Lacaprara\thanksref{addr9}, K.~Lieret\thanksref{addr17}, R.~Maiti\thanksref{addr15}, A.~Martini\thanksref{addr2}, F.~Meier\thanksref{addr4}, F.~Metzner\thanksref{addr14}, M.~Milesi\thanksref{addr21}, S.-H.~Park\thanksref{addr7}, M.~Prim\thanksref{addr20}, C.~Pulvermacher\thanksref{addr14}, M.~Ritter\thanksref{addr17}, Y.~Sato\thanksref{addr7}, C. Schwanda\thanksref{addr15}, W.~Sutcliffe\thanksref{addr20}, U.~Tamponi\thanksref{addr12}, F.~Tenchini\thanksref{addr10}, P.~Urquijo\thanksref{addr21}, L.~Zani\thanksref{addr1}, R.~\v{Z}leb\v{c}\'{i}k\thanksref{addr5}, A.~Zupanc\thanksref{addr16}
}

\thankstext{e1}{e-mail: paul.feichtinger@oeaw.ac.at (corresponding author)}


\institute{Aix Marseille Universit\'{e}, CNRS/IN2P3, CPPM, 13288 Marseille, France \label{addr1}
            \and
            Deutsches Elektronen-Synchrotron, Hamburg, Germany \label{addr2}
            \and
            Dipartimento di Fisica, Universit\`{a} di Pisa, I-56127 Pisa, Italy \label{addr3}
            \and
            Dipartimento di Scienze Fisiche, Universit\`{a} di Napoli Federico II, I-80126 Napoli, Italy\label{addr3b}
            \and
            Duke University, Durham, USA \label{addr4}
            \and
           Faculty of Mathematics and Physics, Charles University, Prague, Czech Republic \label{addr5}
           \and
            Helmholtz AI, Karlsruhe Institute of Technology, 76131, Karlsruhe, Germany \label{addr6}
            \and
            High Energy Accelerator Research Organization (KEK), Tsukuba, Japan \label{addr7}
            \and
           INFN - Sezione di Napoli, I-80126 Napoli, Italy \label{addr8}
           \and
           INFN - Sezione di Padova, Padova, Italy \label{addr9}
           \and
           INFN - Sezione di Pisa, I-56127 Pisa, Italy \label{addr10}
           \and
           INFN - Sezione di Roma Tre, Roma, Italy \label{addr11}
           \and
           INFN - Sezione di Torino, Torino, Italy \label{addr12}
           \and
           INFN - Sezione di Trieste, Trieste, Italy \label{addr13}
           \and
           Institut f\"ur Experimentelle Teilchenphysik, Karlsruher Institut f\"ur Technologie, Karlsruhe, Germany \label{addr14}
           \and
           Institute of High Energy Physics, 1050, Vienna, Austria \label{addr15}
           \and
           Jo\v{z}ef Stefan Institute, Ljubljana, Slovenia \label{addr16}
           \and
           Ludwig Maximilians University, Munich, Germany \label{addr17}
           \and
           Max-Planck-Institut f\"ur Physik, M\"unchen, Germany \label{addr18}
           \and
           Tel Aviv University, Tel Aviv, Israel \label{addr19}
           \and
           University of Bonn, Bonn, Germany \label{addr20}
           \and
           University of Melbourne, Melbourne, Australia \label{addr21}
}

\date{Received: date / Accepted: date}

\maketitle

\begin{abstract}
We present the novel implementation of a non-differentiable metric approximation and a corresponding \linebreak
loss-scheduling aimed at the search for new particles of unknown mass in high energy physics experiments. We call the loss-scheduling, based on the minimisation of a figure-of-merit related function typical of particle physics, a Punzi-loss function, and the neural network that utilises this loss function a Punzi-net. We show that the Punzi-net outperforms standard multivariate analysis techniques and generalises well to mass hypotheses for which it was not trained. This is achieved by training a single classifier that provides a coherent and optimal classification of all signal hypotheses over the whole search space.
Our result constitutes a complementary approach to fully differentiable analyses in particle physics.
We implemented this work using PyTorch and provide users full access to a public repository containing all the codes and a training example.
\end{abstract}

\section{Introduction}
\label{intro}
The standard model (SM) of particle physics is the theoretical framework that describes fundamental interactions and the fundamental constituents of matter. Although successful in predicting phenomena, there is a general consensus that this framework is not a complete description of nature, and new physics (NP) has to exist. 
Searches for NP beyond the SM can be grouped into two main categories: searches for direct production and decays of new, unknown particles; and searches for deviations from the theoretical predictions in precision measurements. When searching for new particles, for example, in a collider experiment, one of the main challenges is correctly reconstructing and identifying the new particles (the signal) and rejecting any (or most) contributions from potential background sources. 
This is a common problem referred to as event classification. A common approach to correctly classify a signal with respect to background uses Monte Carlo (MC) simulation to generate signal- and background-like event distributions.
MC simulation can help find underlying features or patterns in the signal and the background distributions that allow one to disentangle the two (possibly) unambiguously. In the last decade, advanced data analysis methodologies, such as multivariate analysis (MVA) methods, have often improved analysis signal selection power, allowing for more precise analyses, usually performed in a shorter time. 
Typical MVA methods in use in the field of particle physics include, but are not limited to, decision trees,
boosted decision trees (BDTs)~\cite{Keck:2017gsv}, or shallow and deep neural networks (NNs)~\cite{albertsson2019machine}. 
This paper focuses on the implementation of NNs. We propose and describe how to implement a new loss function, called Punzi-loss, based on the so-called Punzi figure-of-merit (FOM)~\cite{Punzi:2003bu}. We henceforth refer to a neural network trained with the Punzi-loss function as a Punzi-net. 
As a benchmark study to test the performance of the Punzi-loss and compare it to other techniques, we consider the search for invisible decays of the hypothetical \zprime boson produced in the reaction $e^+e^- \to \mu^+ \mu^- \zprime$ at the Belle II experiment~\cite{Abe:2010sj,Kou_2019} at the SuperKEKB collider \cite{superkekb}, based on MC simulations.

\section{Neural networks}
\label{sec:ann}
There exist many implementations of neural networks (e.g. convolutional neural networks (CNNs), transformers, etc.) that are used in various applications ranging from image classification in the case of CNNs to natural language processing with transformers. In this work, we focus on a fully connected feed-forward neural network for our experiments. We nonetheless emphasise that the concepts outlined in this work apply to all neural network implementations that use backpropagation.

A neural network comprises a collection of connected neurons. In a fully connected neural network, these constitute a series of layers in which each neuron is connected to all those in both the previous and subsequent layers. Each neuron describes a mathematical function that produces an output dependent on those input connections and a unique bias, defined as
\begin{equation}\label{neuron_output}
 a_{j}^l = \sigma\left(\sum_{k}{w_{jk}^{l}a_{k}^{l-1} + b_{j}^l }\right). 
\end{equation}
Here $w_{jk}^l$ is the weighting of the connection to the $k^{th}$ neuron in the previous ($l-1$) layer, $b_{j}^l$ is the bias and $\sigma$ is the \emph{activation function}. A variety of different activation functions can be applied here, and most have specific traits that may be desirable depending on the application. Commonly used examples include sigmoid,  rectified linear activation (ReLU), or hyperbolic tangent functions. The key requirements are that they are non-linear and have a derivative defined everywhere.

Using Eq.~\ref{neuron_output}, a network of individual neurons is able to map input variables to some desired output. For this to be possible, however, the weight and bias parameters must be optimised. In the implementation we present here, this is done via \emph{supervised training}, whereby training data, $x$, is passed to the network along with the set of corresponding labels, $y$. The actual output of the network,  $f(x)=\hat{y}$, can then be compared with this desired output to measure how well it maps input data. This comparison is quantified by way of a loss function, a commonly used example of which is the Binary Cross Entropy loss,
\begin{equation}\label{eq:bce_loss}
L=-y\ln{\hat{y}}-(1-{y}) \ln(1-\hat{y}),
\end{equation}
where $y \in \{0, 1\}$ and $\hat{y}$ $\in [0,1]$. With this measure of the error, the training process becomes a minimisation problem: what weights and biases will minimise the loss and therefore provide the most effective network? This is solved by employing a method such as \emph{gradient descent}, by which the parameters can be iteratively adjusted in the direction opposite that of the loss function's gradient,
\begin{equation}\label{weight_update}
 w_{n+1} = w_{n} - \eta\frac{\delta L}{\delta w_n} ~\text{and}
\end{equation}
\begin{equation}\label{bias_update}
 b_{n+1} = b_{n} - \eta\frac{\delta L}{\delta b_n},
\end{equation}
where $\eta$ is the \emph{learning rate}, the step size by which the parameters are adjusted at each iteration of the learning process. Each of these iteration steps constitutes a complete pass through a randomly sampled \emph{batch} from the full training data set, a pass through the entirety of which is referred to as an \emph{epoch}. The derivatives $\frac{\delta L}{\delta w_j}$ and $\frac{\delta L}{\delta b_j}$ are calculated through use of the \emph{backpropagation} algorithm. This starts from the final layer and utilises the chain rule to incrementally calculate all derivatives through one full backward pass to the first layer. The key is carefully selecting a loss function whose minimum solves the given task while remaining differentiable across all possible neural network outputs.

\section{Figure of merit}
As highlighted in Section~\ref{intro}, one of the main challenges when performing a precision test of the SM, or in the search for NP, is the fact that some background processes may mimic the signal and therefore contaminate the results. In the search for a new particle, for example, one is often performing a counting experiment which is described by the Poisson distribution. As discussed in~\cite{Punzi:2003bu}, the number of events $n$ in a counting experiment in the case of a background ($B$) only hypothesis ($H_B$), and in the case of a signal ($S$) in the presence of the same background ($H_{S+B}$) follows the Poisson distributions
\begin{equation}\label{H0}
 p(n \mid H_B) = \frac{B^n e^{-B}}{n!}
\end{equation}
and
\begin{equation}\label{Hs}
 p(n \mid H_{S+B}) = \frac{(S+B)^n e^{-(S+B)}}{n!}.
\end{equation}

When MC simulations for both the signal and the background are available, it is possible to identify quantities or features in the data to separate and classify them correctly by applying specific selection criteria. This would eventually enable one to choose between the (null) background only and the signal plus background hypotheses. In general, however, applying some selection criteria to reduce the background contamination will also remove some of the signal.
It is, therefore, fundamental to define some additional criteria that would indicate the best balance between reducing the background without compromising the signal. 
This is done via the implementation of a FOM. One can define $S(t)$ and $B(t)$ as the number of signal and background events that pass some selection criteria (e.g. particles having a momentum or energy larger than a specified threshold $t$). 
In that case, standard FOMs used in particle physics are:
\begin{equation}\label{FOM1}
 FOM= \frac{S(t)}{\sqrt{B(t)}} ~\text{and}
 \end{equation}
\begin{equation}\label{FOM2}
 FOM= \frac{S(t)}{\sqrt{S(t)+B(t)}}.
\end{equation}

Neither of the above is usable in the search for new particles since the number of expected signal events depends on the cross-section of the process, and this is not known \textit{a priori}. An alternative FOM for this specific case was proposed in~\cite{Punzi:2003bu}, often referred to as the Punzi FOM after the author, and is now in widespread use. The Punzi FOM to maximise is the inverse of the minimum detectable cross-section $\sigma_{\min}$, which defines a sensitivity region for which the experiment will certainly give conclusive results: it either will be excluded, or a discovery will be claimed.
An analytic formula for $\sigma_{\min}$ is given by
\begin{equation}
\sigma_{\min }(t)= \frac{\frac{b^{2}}{2}+a \sqrt{B(t)}+\frac{b}{2} \sqrt{b^{2}+4 a \sqrt{B(t)}+4 B(t)}}{\epsilon(t) \cdot L},
\label{eq:punzi_sens}
\end{equation}
where $L$ is the target luminosity, $\epsilon(t)$ is the signal efficiency and $B(t)$ is the number of background events after the selection defined by $t$.
The constants $a$ and $b$ are the number of sigmas corresponding to one-sided Gaussian tests at some predefined significance level, $\alpha$ and $\beta$.
Here $\alpha$ is the probability of rejecting $H_B$ when it is true (type I error), and $\beta$ is the probability of not rejecting $H_B$ when instead $H_{S+B}$ is true (type II error).
Since we are interested in cases where the signal hypothesis depends on a free parameter, $\beta$ will change with this parameter.
The sensitivity region for a given experiment is obtained for the parameter space that fulfils $1-\beta > CL$, where $CL$ is the confidence level for setting limits in case of no discovery.
So for example when choosing $\alpha$ to correspond to a significance of $5\sigma$ and a desired confidence level of \SI{90}{\percent}, $a$ and $b$ would be set to \num{5} and \num{1.28}.

\section{Punzi-loss}
\label{sec:punzi}
We propose here a quantity approximating the Punzi FOM, appropriate for optimising neural networks for physics selections.

This loss function is based on the equation for the Punzi sensitivity region (Eq.~\ref{eq:punzi_sens}).
However, Eq.~\ref{eq:punzi_sens} can not be used directly because the number of background events $B$ and the signal efficiency $\epsilon$ are discrete functions of the network parameters for any given fixed cut on the classifier output, whereas the loss function must be differentiable.
We can build a differentiable function by replacing the fixed cut on the output with a sum over all events, weighted with the respective value of the output.
If events classified as signal cluster around an output of 1 and events classified as background at 0, this quantity will closely approximate the original function.
In Eq.~\ref{eq:punzi_sens} this weighting can be captured by performing the replacements

 \begin{align}
\label{eq:eff_replace}
  \epsilon(t) & \rightarrow \epsilon(\boldsymbol{w}, \boldsymbol{b}) = \sum_{\boldsymbol{x}} \frac{y_{i}\cdot \hat{y}_{i}(\boldsymbol{w}, \boldsymbol{b}) \cdot s_{\mathrm{sig}}}{N_{\mathrm{gen}}} \quad \text{and} \\[.2cm]
\label{eq:bkg_replace}
  B(t)        & \rightarrow B(\boldsymbol{w}, \boldsymbol{b}) = \sum_{\boldsymbol{x}} (1 - y_{i})\cdot \hat{y}_{i}(\boldsymbol{w}, \boldsymbol{b}) \cdot s_{\mathrm{bkg}}^{\,i},
 \end{align}%
where the sum is over all training inputs $\boldsymbol{x}$ and the index $i$ denotes the $i^{\text{th}}$ training event.
The collection of weights and biases that constitute the free parameters of the network are denoted as $\boldsymbol{w}$ and $\boldsymbol{b}$.
$N_{\mathrm{gen}}$ is the total number of generated signal events, $s_{\mathrm{sig}}$ is a scale factor for the signal and 
$s_{\mathrm{bkg}}^{\,i}$ is a scale factor for the background, which can include a weight factor to scale the luminosity for the individual simulated background samples to the target luminosity.
The scale factors can also include correction factors such as trigger efficiencies and should account for the sample size when only a subset of the generated data is used to compute the loss.
A similar approach of building a differentiable metric based on a FOM was taken by Elwood and Krücker \cite{elwood2018direct}, with a loss function based on the discovery significance.

The Punzi-loss function is given by the arithmetic mean of this continuous Punzi sensitivity calculated for all signal hypotheses ($m_{\zprime}$) that are used in training,
\begin{equation}
 \label{eq:punzi_loss}
 C_{\text{Punzi}}= \frac{1}{N_{\zprime}}\ \mathlarger{\sum_{m_{\zprime}}}\ \sigma_{\min }(\boldsymbol{w}, \boldsymbol{b}),
\end{equation}
with $N_{\zprime}$ being the total number of hypotheses that were considered. Here we present an implementation in which all mass hypotheses are treated equally since they have equal weights in the calculation. However, one could introduce some weightings in the case of an analysis where the hypotheses do not have flat priors. 
Note that this loss function can no longer be calculated using single training events but is instead based on a set of training data.

To test the Punzi-loss function, we implemented a simple fully-connected network in \texttt{PyTorch} \cite{pytorch} with four input neurons, one output neuron, and two hidden layers with 8 and 4 neurons, respectively.
The size of the net was determined empirically to give good results while keeping the network relatively small.\footnote{The network size and architecture is not relevant for our approach.}

\section{Training strategy}
\label{sec:training}
For the Punzi-loss training to converge, we found that the parameters of the network should already be initialised in a way that defines some separation between signal and background (similar to the loss scheduling scheme described in \cite{loss_scheduling}). This can be achieved by pretraining the network using a conventional loss function and subsequently fine-tuning this through the use of the Punzi-loss function.

\begin{figure}[ht]
 \centering
\includegraphics[width=\linewidth]{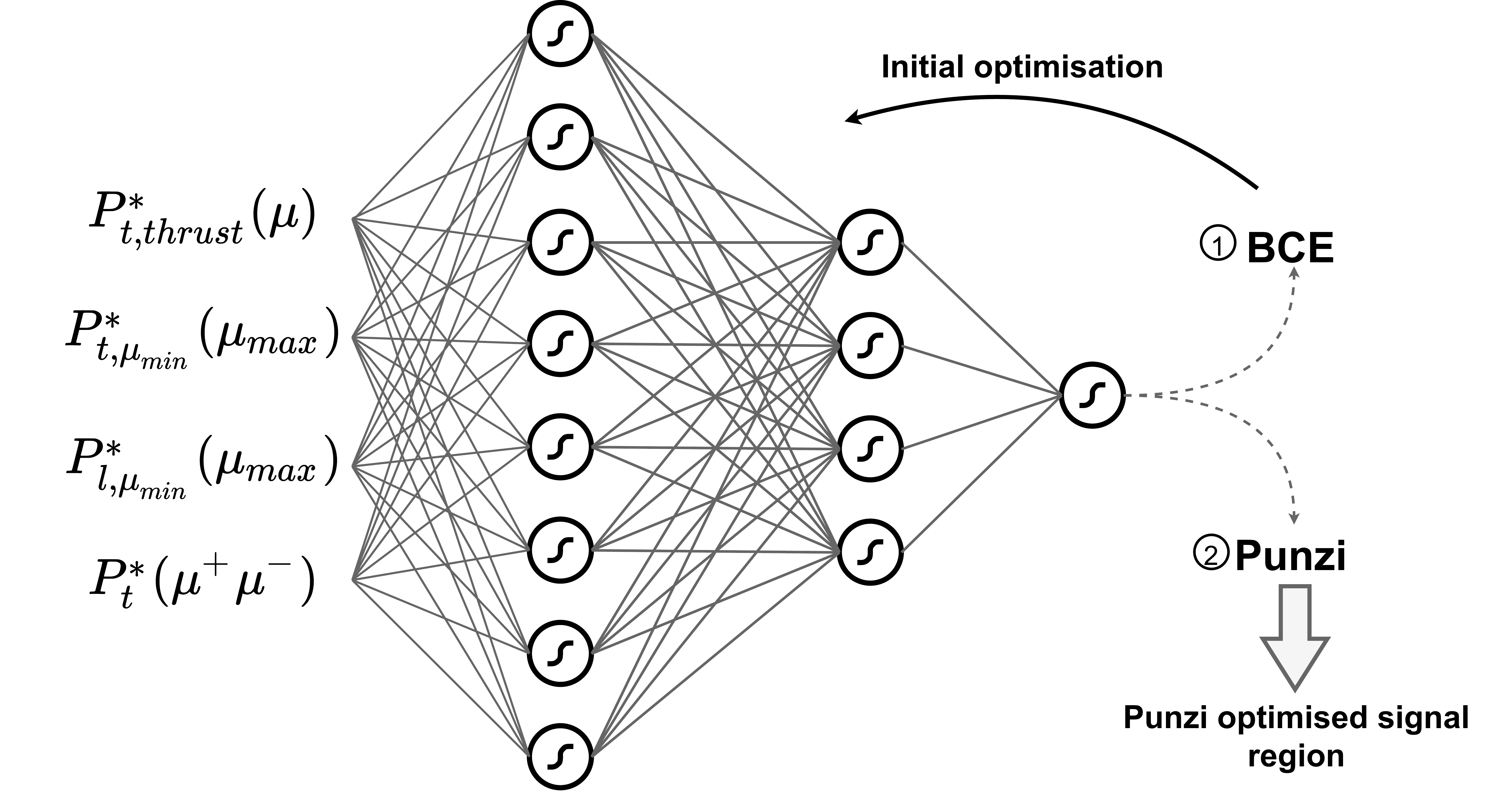}
 \caption{An outline of the network architecture. The first training with the BCE loss function was used to set the weights and biases of the net for the second training with the custom loss function based on the Punzi FOM.}
 \label{fig:net_sketch}
\end{figure}

For the activation function of the neurons in the hidden layers, a hyperbolic tangent is used while the output neuron uses a sigmoid function.
Before training, the input variables were scaled to lie between 0 and 1, and the network parameters were randomly initialised. A weighted binary cross-entropy (BCE) loss function was used for the pretraining. A weighting was attributed to the signal events such that their weighted sum was equal to the weighted sum of all background events. 
An outline of the network architecture is given in Figure~\ref{fig:net_sketch}.

Initially, using the BCE loss function, the network was trained with a batch size of \num{2048} and a learning rate (LR) of 1. When the loss did not decrease for \num{10} epochs, the LR was reduced by a factor of 0.5. The pretraining was stopped after \num{200} epochs. Training is then continued using the Punzi-loss function with $a=3$ and $b=1.28$.
Here we used a learning rate of \num{0.0001} and again reduced it upon plateauing. This training was stopped after \num{1000} epochs. The gradient descent algorithm was used for optimisation for both of these trainings. 
All hyperparameters were optimised to give the best results for the training methods.
One particularly important hyperparameter is the batch size, the variation of which presents some unique aspects of the Punzi-loss function that must be considered.

Due to the nature of the Punzi-loss function concerning the optimisation for a desired luminosity, utilising training data in excess of this requires the addition of weightings in the loss calculation. The background data used for training contained \SI{1000}{\invfb}, \SI{450}{\invfb} and \SI{3000}{\invfb} worth of events from three main background processes; however, in this study, we wish to optimise the classifier for just \SI{50}{\invfb} of real-world data. Naturally, it is preferred that all background data is utilised, and thus we introduce a background scaling factors of \num{0.05}, \num{0.111} and \num{0.0167} respectively. Additionally, dividing the training data into batches brings about the additional requirement of multiplying both $\epsilon(\boldsymbol{w}, \boldsymbol{b})$ and $B(\boldsymbol{w}, \boldsymbol{b})$ by the number of batches used.

\begin{figure}[!ht]
 \centering
 \includegraphics[width=1\linewidth,trim={1.5cm 0cm 1.6cm 1.5cm},clip]{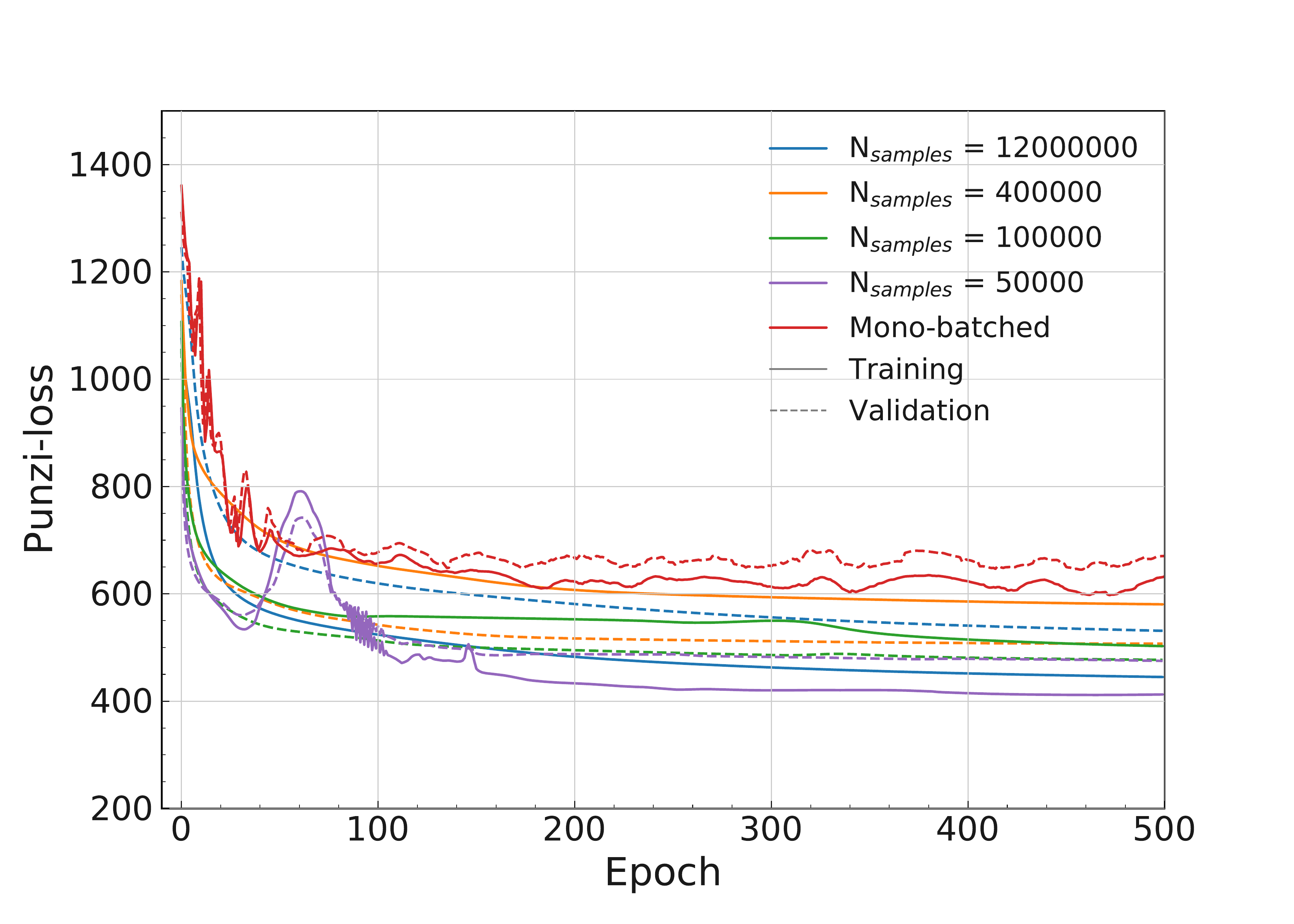}
 \caption{Evolution of Punzi-loss during training with batch sizes of \num{5e4}, \num{1e5}, \num{4e5} and \num{1.2e6} and mono-batched. }
 \label{fig:loss_evoluton}
\end{figure}

Fig. \ref{fig:loss_evoluton} shows the evolution of loss during training with batch sizes of \num{5e4}, \num{1e5}, \num{4e5} and \num{1.2e6}, and also mono-batched (where the whole data set is passed as a single batch). These correspond to batch sizes of between roughly \num{0.5}$\%$ and \num{13}$\%$ of the total dataset. A batch size of \num{1e5} was chosen for the following experiment. We note that small batch sizes bring a degree of instability to the loss, as can be seen in the line representing a batch size of \num{5e4} in fig. \ref{fig:loss_evoluton}. It was found that batches smaller than those shown in Fig.~\ref{fig:loss_evoluton} led to increasing loss values over the training, with a batch size of \num{1e4} leading to training regularly failing with the Punzi-loss increasing and plateauing at a value above the initial loss. This can be understood as a result of the limited number of signal events present in any given batch of small size, leading to large statistical fluctuations in the calculated loss values. This lower limit is, of course, study dependent. Similarly, we note large instability in the mono-batched case. 
The batching introduces an additional stochastic component during training, making the training more robust and helping the algorithm escape local minima.

\section{Results}
\label{sec:results}
In this section, we present the results of utilising a Punzi-net in a search for $e^+e^- \to \mu^+ \mu^- \zprime$ signals amongst various common backgrounds found in $e^+e^-$ collider experiments. At the Belle II experiment, this search was performed with the commissioning data for the specific case of invisible decays of the $\zprime$ boson~\cite{PhysRevLett.124.141801}, a final state in which only the two muons produced by the electron-positron annihilation can be reconstructed. Therefore, all information about the production and decay of the $\zprime$ boson is to be inferred by the two-muon system.
The signal events are generated with \texttt{MadGraph 5} 
\cite{madgraph5}
for a range of candidate \zprime masses, spanning \SIrange{0.1}{8.9}{\gevcc} in steps of \SI{0.1}{\gevcc} with \num{20000} events produced at each. 
Additionally, MC samples for the background process $e^+e^- \to e^+ e^- \mu^+ \mu^-$, $e^+e^- \to \tau^+ \tau^-$ and $e^+e^- \to \mu^+ \mu^- (\gamma)$ corresponding to \SI{1000}{\invfb}, \SI{450}{\invfb} and \SI{3000}{\invfb} respectively were used, since these can mimic the signal.
The simulation and reconstruction of the events were done using \texttt{GEANT4} \cite{GEANT4}, and the Belle II Analysis Software Framework \cite{Kuhr2018}.
The analysis is carried out via the search for a peak in the distribution of the squared mass recoiling against the two-muon system. An excess of entries beyond that of the expected background at a given mass would indicate the presence of such a \zprime particle of that mass. This distribution is divided into (potentially overlapping) bins with bin widths corresponding to $\pm 2 \sigma$ of the fitted \zprime signal distributions.

During both the initial BCE and subsequent Punzi-loss training, only every second generated \zprime mass was used. For the calculation of $\sigma_{\text{min}}$ in Eq.~\ref{eq:punzi_loss} only signal and background events that lie within the respective $\pm 2 \sigma$ mass windows are considered, using only signal events that were generated for the corresponding mass.
Thus, events that are not contained in any of the mass windows of the used signal samples are not taken for the training. 
This results in a data set of approximately 9 million total events, of which $\sim$2.5\% are signal and the rest background. This is then split by using 80\% of the events for training and the remaining 20\% for validation. The unused signal hypotheses are utilised for validation and to check the trained networks ability to generalise to signals unseen in training. The network was trained with four carefully selected features related to the event kinematics that showed a good discrimination power when using a boosted decision tree classifier. These features are described in Tab.~\ref{tab:features}. A more detailed description of these features and the analysis can be found in \cite{thesis2021}.

\begin{table}
\centering
\begin{tabular}{ll}
 \hline
 variable                      & description                                                     \\
 \hline
 \multirow{2}{*}{\ptThrust}    & The transverse momentum component \\
                                 & of the muons with respect to the thrust axis.                                \\
 \multirow{3}{*}{\ptMaxWrtMin} & The transverse momentum component of the  \\
                                &  higher energetic muon with respect to  \\
                                &  the lower energetic muon.       \\
 \multirow{3}{*}{\plMaxWrtMin} & The longitudinal momentum component of the   \\
                                & higher energetic muon with respect to the lower     \\
                                & energetic muon.    \\
  \multirow{1}{*}{\ptMumu}    & The transverse momentum of the dimuon system.  \\
  \hline
\end{tabular}
\caption{The most important features found after training BDTs with many variables. All variables are computed in the centre-of-mass system of the $e^+e^-$ collisions. These features are used for training the NN.}
\label{tab:features}
\end{table}

\begin{figure}[!ht]
 \centering
 \includegraphics[width=1\linewidth,trim={0.5cm 0cm 1.5cm 0.5cm},clip]{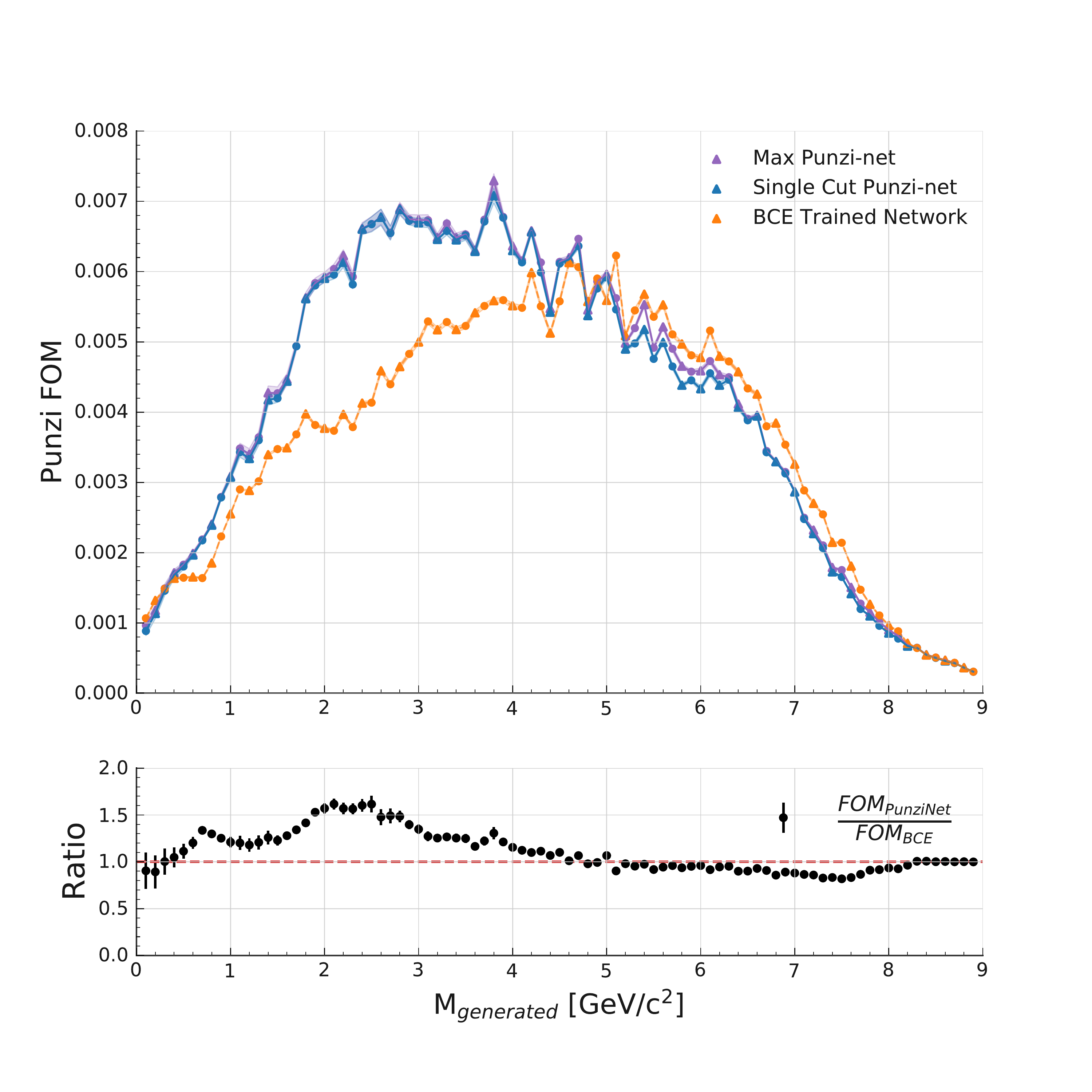}
 \caption{The average maximum Punzi FOM achievable in each bin across range of generated \zprime signals, with standard error spread taken from 10 independently trained networks. Triangles indicates those masses which were left out of training while circles indicates those used. }
 \label{fig:Punzi_FOM}
\end{figure}

The resulting maximum achievable Punzi FOM spanning the range of generated \zprime signals is shown in Fig.~\ref{fig:Punzi_FOM}. Included in this figure are the Punzi-net along with the BCE pretrained network. These values are calculated using the background data contained within the $\pm 2 \sigma$ bin around each generated mass point. The maximum achievable Punzi FOM in each bin is found using the cut to the network output that provides the highest FOM for that respective bin. In addition to this, the resulting Punzi FOM after applying a single cut value to the output of the Punzi-net across the full recoil mass spectrum is shown. The plot shows the average result found over ten independently trained networks, along with the associated standard error. This serves to demonstrate that not only can the Punzi-loss function produce better FOMs, but can do so consistently. The Punzi-loss function shows greater effectiveness through 
the lower half of the recoil mass spectrum, providing clear improvements to the FOM below approximately 5 {\gevcc}. For mass hypotheses above this point, there is some slight degradation of the maximum achievable FOM with the Punzi-net. It is important to note that the single cut applied to the Punzi-net output can still provide a FOM near to that of the maximum achievable with the BCE trained network in this region.  

This means that even when compared to an optimal varied cut applied to the BCE network output, interpolated over the recoil mass spectrum, the Punzi-net provides comparable or even improved results. As discussed previously, this cut interpolation can lead to discontinuities in the final recoil mass distribution. So the ability to achieve comparable results with a single cut to the Punzi-net output is much preferable, meaning that even in the higher recoil mass region where the BCE network appears to outperform the Punzi-net, it may not be a preferable method due to the need for cut interpolation.  

The generated \zprime masses used for training the network are shown with circles, and those not used in training are shown with triangles. 
The figures show little to no difference in the network's ability between these training and validation masses, indicating that the model generalises well to unseen signals. In the region between approximately 4.5 {\gevcc} and 5.5 {\gevcc} some dependence on whether or not a mass was used in training does appear. This could be combated by generating a larger set of \zprime signals covering more mass points in that region. 

\begin{figure}[ht]
 \centering
 \includegraphics[width=1\linewidth,trim={0.5cm 0cm 0.5cm 0.5cm},clip]{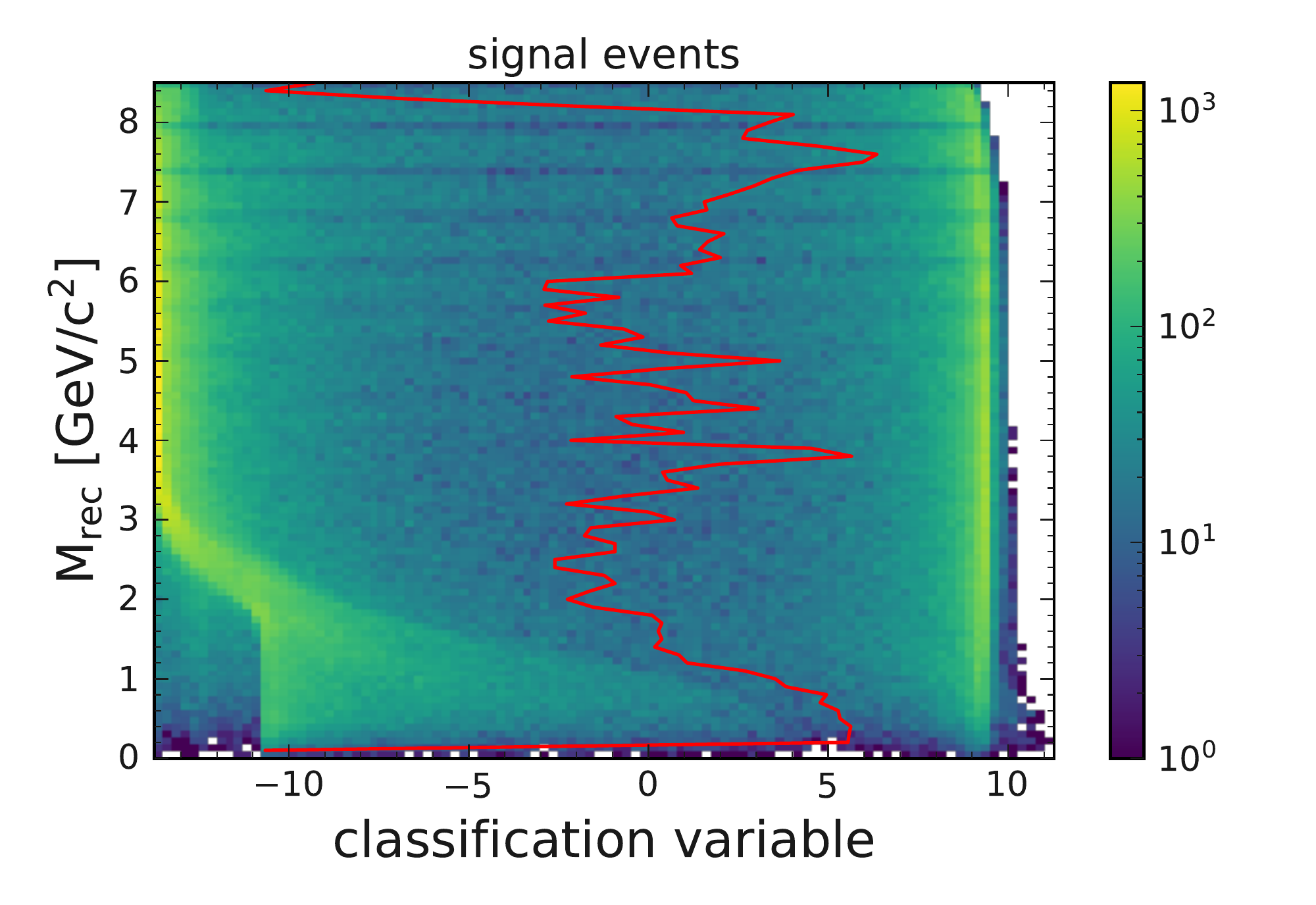}
 \caption{The output distribution of all signal events using the Punzi-loss trained NN, overlaid with the the optimal decision threshold for each signal hypothesis. The classification variable shows the NN output before applying the last sigmoid function in order to better see the separation. The optimal cut value can be replaced by a uniform cut without any significant difference in the resulting selection.}
 \label{fig:signal_output}
\end{figure}

\begin{figure}[ht]
 \centering
 \includegraphics[width=1\linewidth,trim={0.5cm 0cm 0.5cm 0.5cm},clip]{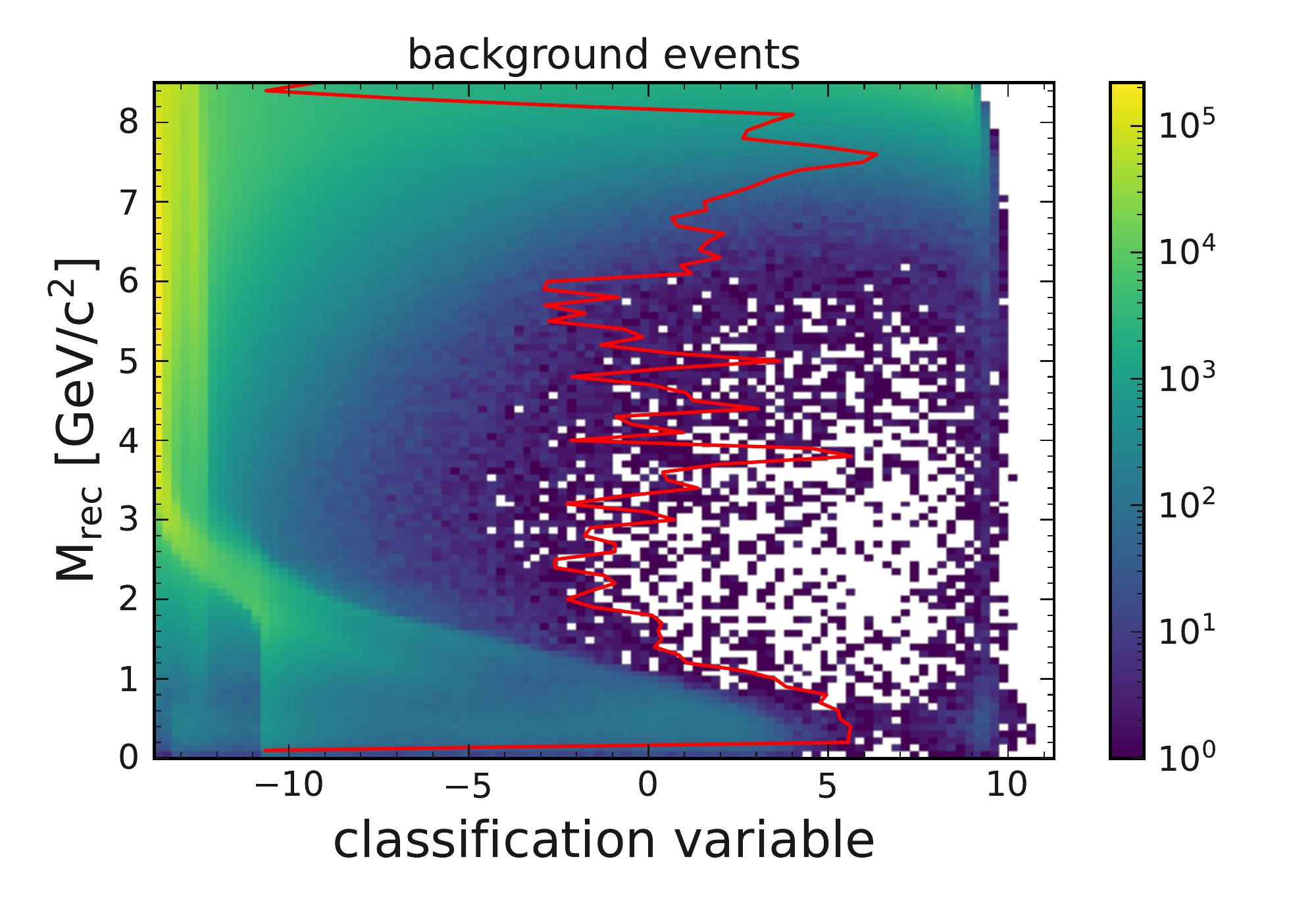}
 \caption{The output distribution of all background events using the Punzi-net, overlaid with the optimal decision threshold for each signal hypothesis.}
 \label{fig:background_output}
\end{figure}

Fig.~\ref{fig:signal_output} shows the output of the Punzi-trained network for all signal events and Fig.~\ref{fig:background_output} shows the same for all background events.
Here the variable on the x-axis shows the NN output before applying the last sigmoid activation function to resolve the distribution of events better.
The y-axis corresponds to the reconstructed recoil mass (M$_{\text{rec}}$), which discriminates between the different signal hypotheses.
The classified signal and background events are separated into two clusters, corresponding to an output of 0 and 1.
The overlaid line shows the cut value that would give the maximum achievable Punzi FOM for each \zprime mass.
The line separates the two clusters, showing that the training using the approximations in Eq.~\ref{eq:eff_replace} and \ref{eq:bkg_replace} worked as expected.

The events are separated so that when only the events classified as signal are selected (for example, by applying a cut at a NN output of 0.5), this gives the optimal Punzi FOM for the whole mass range.
This is a significant advantage for an analysis since no additional interpolation between output values is required, which can introduce discontinuities in the final recoil mass distribution. %
Additionally, since the selection generalises to all signal hypotheses, it gives also the best possible FOM for a signal in-between trained masses, which would otherwise have non-optimal results.

\begin{figure}[ht]
 \centering
 \includegraphics[width=1\linewidth,trim={0.5cm 0cm 0.5cm 0.5cm},clip]{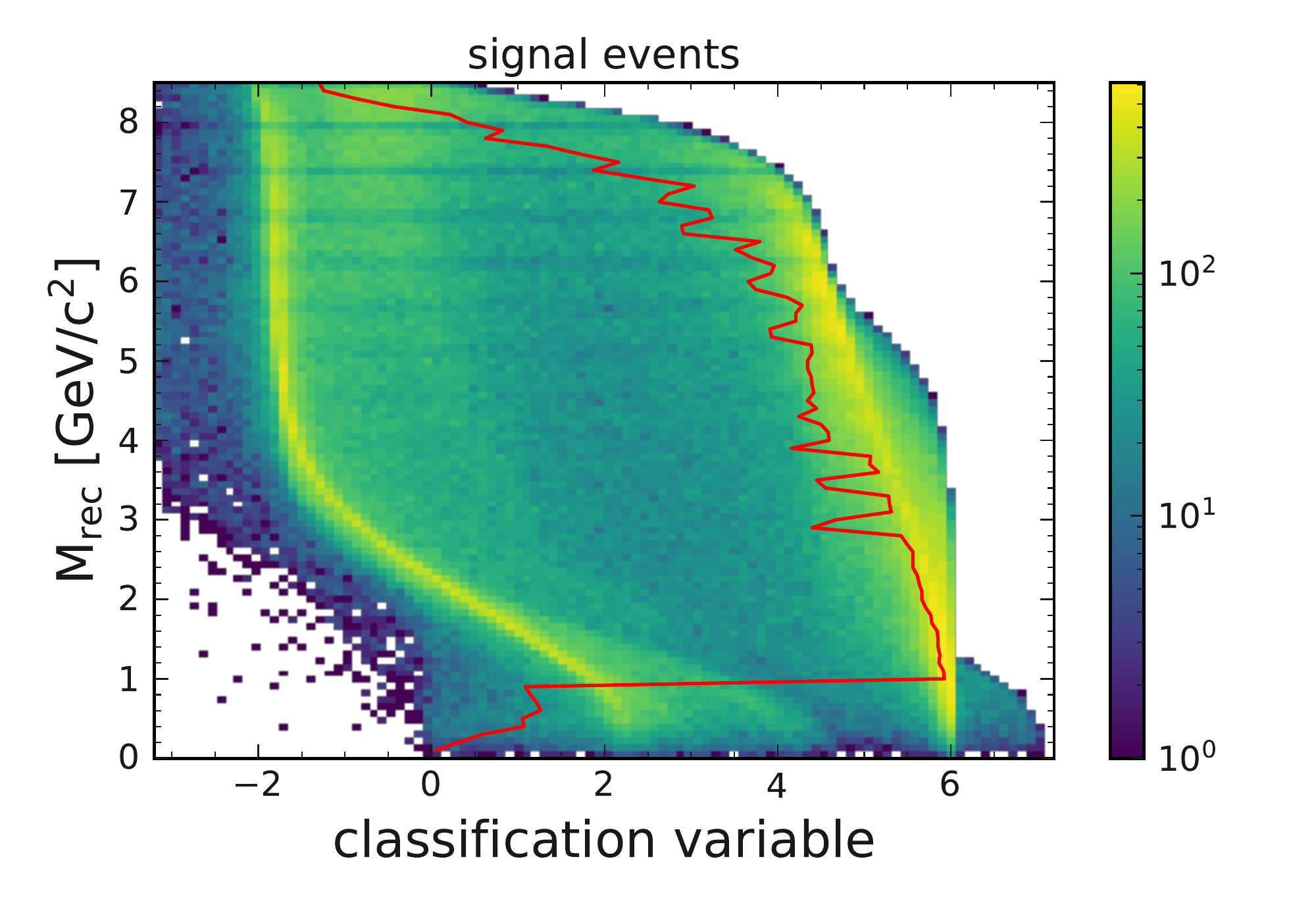}
 \caption{The output distribution of all signal events after the BCE pretraining, overlaid with the optimal decision threshold for each signal hypothesis. The optimal cut value varies significantly across the mass spectrum.}
 \label{fig:signal_output_BCE}
\end{figure}

\begin{figure}[ht]
 \centering
 \includegraphics[width=1\linewidth,trim={0.5cm 0cm 0.5cm 0.5cm},clip]{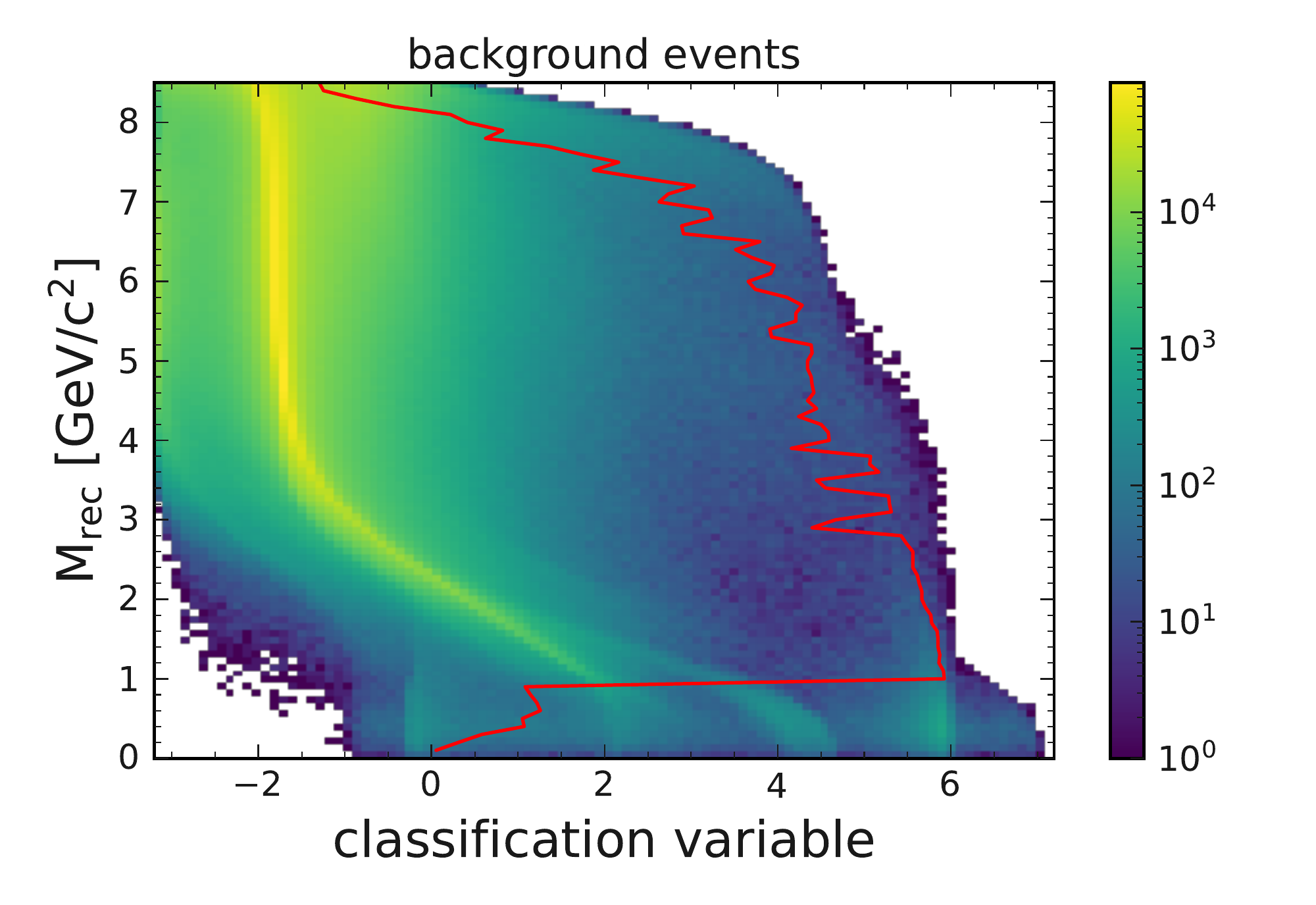}
 \caption{The output distribution of all background events after the BCE pretraining, overlaid with the optimal decision threshold for each signal signal hypothesis.}
 \label{fig:background_output_BCE}
\end{figure}

For comparison we also show the output distribution of the BCE pretrained network in Fig.~\ref{fig:signal_output_BCE} for the signal and Fig.~\ref{fig:background_output_BCE} for the background.
Again, the optimal cut that gives the highest Punzi FOM at each mass hypothesis is shown.
While a separation between signal and background is also achieved here, the division is not as pronounced as with the Punzi-net and the best cut value varies significantly with the mass.

\begin{figure}[ht]
 \centering
\includegraphics[width=\linewidth]{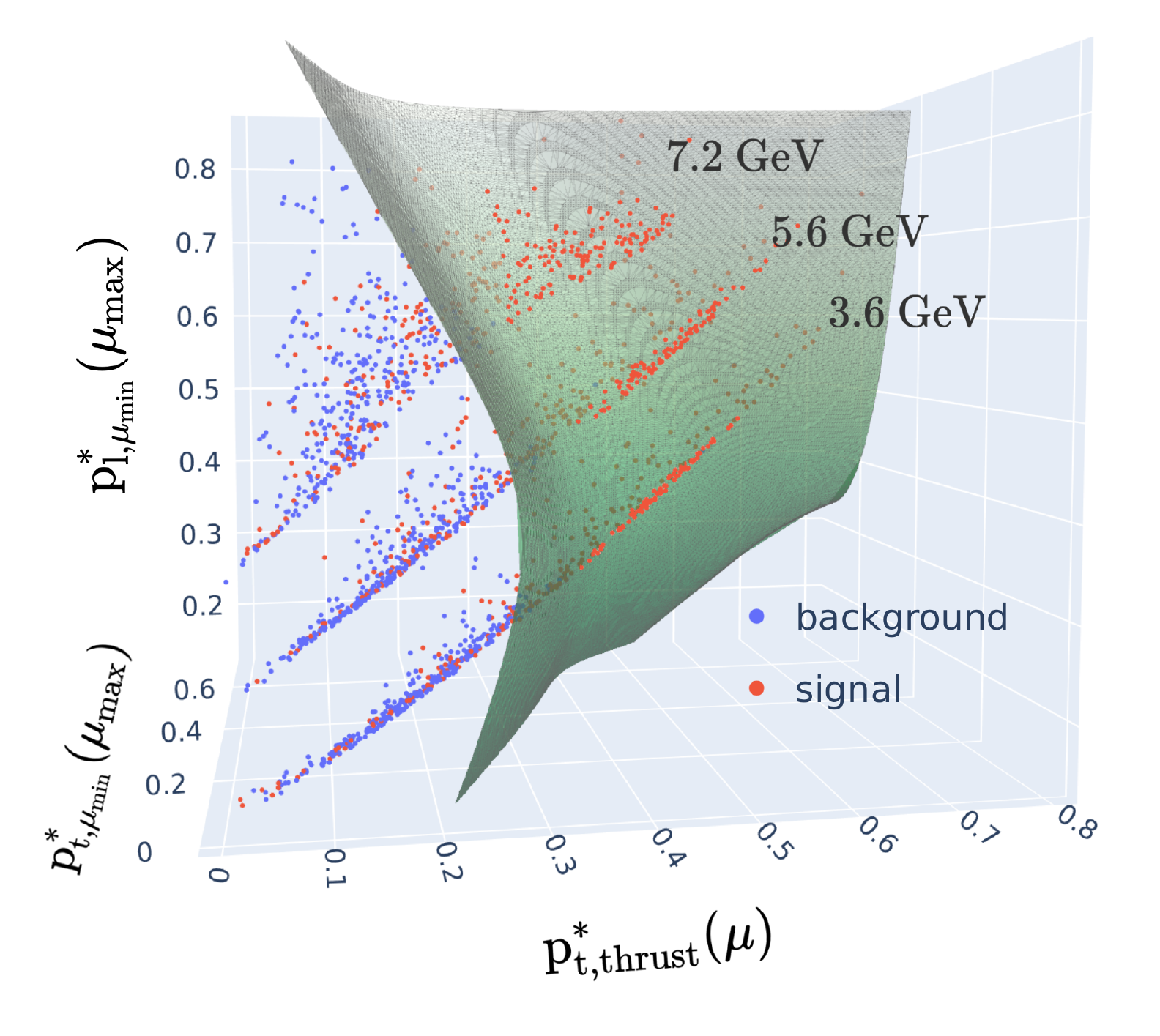}
 \caption{A 3D scatter plot showing the input space of the NN with $\ptMumu$ fixed around $\SI{2.2}{\gevc}$.
  The separation boundary defined by the final selection (green sheet) separates the planes corresponding to different recoil masses in a way that optimises the selection for all signal hypotheses.}e
\label{fig:net_3d}
\end{figure}

An understanding of why the model successfully generalises, and one network can be utilised for the full squared recoil mass spectrum, can be inferred from Fig. \ref{fig:net_3d}, which shows a 3D scatter plot of the \ptThrust, \ptMaxWrtMin and
\plMaxWrtMin variables (after being normalised to values between 0 and 1) for three of the mass bins at a region of
\ptMumu = $\SI[separate-uncertainty=true]{2.2 \pm 0.5}{\gevc}$.
The green plane is the chosen signal/background classification boundary obtained with a single cut. One can see the masses describing three respective planes in the parameter space which occupy distinct regions. This partitioning allows the network to adapt between the different mass regions and so negates any need for multiple classifiers for different regions. 

\section{Conclusions}
\label{sec:conclusions}
In this work, we have demonstrated that it is possible to implement a non-differentiable metric approximation and a corresponding loss-scheduling, combining the approach of particle physics and that of machine learning.
Our proposed method applies to the search for new particles with unknown parameters in high energy physics experiments.

We designed a new loss function directly related to the Punzi figure-of-merit, intended to be calculated on a set of training events at once.
Training instabilities could be solved by a batched training that helps in the algorithm's convergence.
We showed that this loss function can be used to achieve an optimal selection for all signal hypotheses with a single cut on the classifier output, also achieving overall better performance than standard methods.
The main advantage of this method is that it simplifies the analysis since it does not require any further optimisation of the selection or training of multiple classifiers for subsets of signal hypotheses.
We implemented the Punzi-loss in the training of a simple neural network and made the code publicly available~\cite{public-code}. 
However, the method is general and not restricted to the use of the presented architecture.

A universal approach to this problem would be to construct a fully differentiable analysis pipeline that can optimise any utility function, 
which is an active area of research \cite{ml_experimental_design}.
Such analysis frameworks can also take into account systematic effects during the optimisation of the signal selection \cite{inferno,sys_inference}.
Another interesting approach to incorporate systematic effects is to introduce an adversarial discriminator in addition to a classifier.
This provides a handle for robust inference by learning a pivotal quantity - a predictive function that is insensitive against the unknown values of the nuisance parameters that model the systematic effects. \cite{louppe2017pivot}


\begin{acknowledgements}
The authors would like to thank the Belle II Collaboration and Belle II software group for useful discussions and suggestions on how to improve this work. P. Feichtinger, H. Haigh and G. Inguglia would like to acknowledge funding received under the Horizon 2020 framework of the European Research Council, ERC StG Nr. 947006 \textit{InterLeptons}, and under the FWF standalone framework with grant Nr. P31361 \textit{Searches for dark matter and dark forces at Belle II}.
J. Kahn's work is supported by the Helmholtz Association Initiative and Networking Fund under the Helmholtz AI platform grant. F. Meier work is supported by the US Department of Energy.

\end{acknowledgements}

\bibliographystyle{spphys}       

\bibliography{punzi-loss.bib}   

\end{document}